\documentclass[aps,prl,groupedaddress,twocolumn,showpacs,preprintnumbers,amsmath,amssymb,floatfix,superscriptaddress]{revtex4-1}
\usepackage{graphicx,color}
\usepackage{dcolumn}
\usepackage{bm}
\usepackage{color}
\usepackage{amsmath}
\usepackage{multirow}

\newcommand{\La}{LaAlO$_3$}
\newcommand{\Sr}{SrTiO$_3$}

\begin{document}

\title{\boldmath The Vortex Signature of Discrete Ferromagnetic Dipoles\\at the LaAlO$_3$/SrTiO$_3$ Interface}

\author{A.P. Petrovi\'c}
\affiliation{Division of Physics and Applied Physics, Nanyang Technological University, 637371 Singapore}
\author{A. Par\'e}
\affiliation{Division of Physics and Applied Physics, Nanyang Technological University, 637371 Singapore}
\author{T.R. Paudel}
\affiliation{Department of Physics and Astronomy, Nebraska Center for Materials and Nanoscience, University of Nebraska Lincoln, Nebraska 68588-0299, USA}
\author{K. Lee}
\affiliation{Division of Physics and Applied Physics, Nanyang Technological University, 637371 Singapore}
\affiliation{Cavendish Laboratory, University of Cambridge, Cambridge CB3 0HE, United Kingdom}
\author{S. Holmes} 
\affiliation{Toshiba Research Europe Ltd., Cambridge Research Laboratory, 208 Cambridge Science Park, Milton Road, Cambridge CB4 0GZ, United Kingdom}
\author{C.H.W. Barnes}
\affiliation{Cavendish Laboratory, University of Cambridge, Cambridge CB3 0HE, United Kingdom}
\author{A. David}
\affiliation{Division of Physics and Applied Physics, Nanyang Technological University, 637371 Singapore}
\author{T. Wu}
\affiliation{Division of Physics and Applied Physics, Nanyang Technological University, 637371 Singapore}
\author{E.Y. Tsymbal}
\affiliation{Department of Physics and Astronomy, Nebraska Center for Materials and Nanoscience, University of Nebraska Lincoln, Nebraska 68588-0299, USA}
\author{C. Panagopoulos} 
\affiliation{Division of Physics and Applied Physics, Nanyang Technological University, 637371 Singapore}
\affiliation{Cavendish Laboratory, University of Cambridge, Cambridge CB3 0HE, United Kingdom}

\date{\today}

\begin{abstract}
A hysteretic in-plane magnetoresistance develops below the superconducting transition of LaAlO$_3$/SrTiO$_3$ interfaces for $\left|H_{/\!/}\right|<$~0.15~T, independently of the carrier density or oxygen annealing.  We show that this hysteresis arises from vortex depinning within a thin superconducting layer, in which the vortices are created by discrete ferromagnetic dipoles located solely above the layer.  We find no evidence for finite-momentum pairing or bulk magnetism and hence conclude that ferromagnetism is strictly confined to the interface, where it competes with superconductivity.  
\end{abstract}

\let\thefootnote\relax\footnotetext{\scriptsize{{$^\dagger$}Correspondence and requests for materials should be addressed to A.P.P. (email: appetrovic@ntu.edu.sg) or C.P. (email: christos@ntu.edu.sg)}}


\maketitle

The emergence of ferromagnetic order in a material breaks time reversal symmetry and is hence detrimental to conventional spin-singlet superconductivity.  Since the vast majority of superconductors discovered to date exhibit spin-singlet $s$ or $d$-wave pairing, the coexistence of superconductivity (SC) and ferromagnetism (FM) at the {\La}/{\Sr} interface has proved difficult to reconcile~\cite{Li-2011,Dikin-2011,Bert-2011}.  Bypassing the paramagnetic limit via spin-triplet pairing presents a straightforward solution to the puzzle; however the observation of $s$-wave gaps in doped {\Sr}~\cite{Binnig-1980} and {\La}/{\Sr}~\cite{Richter-2013} together with the loss of inversion symmetry at the interface are unsupportive of such a scenario.  All other mechanisms facilitating SC and FM phase coexistence require a real-space modulation of the SC order parameter, either by creating a spontaneous vortex phase~\cite{Bernhard-2000}, forming finite-momentum electron pairs analogous to a Fulde-Ferrell-Larkin-Ovchinnikov (FFLO) state~\cite{Michaeli-2012}, or spatially dispersing FM and SC over macroscopic lengthscales~\cite{Bert-2011}.  

The nature of the interaction between FM and SC is not the only contentious issue in {\La}/{\Sr}, since the origin and location of the FM moments are also subject to intense discussion.  Purely FM behaviour has only been observed following growth and/or annealing at high O$_2$ pressures~\cite{Brinkman-2007}; conversely, there is strong theoretical support for O$^{2-}$ vacancies~\cite{Pavlenko-2012,*Pavlenko-2012-2} or polar distortions~\cite{Banerjee-2013} causing FM.  Experiments have linked FM with low~\cite{Lee-2011} or high carrier densities~\cite{Moetakef-2012}, describing it as both an interfacial~\cite{Lee-2013} and bulk~\cite{Moetakef-2012} phenomenon. A study of FM and SC in both stoichiometric and O$_2$-deficient interfaces with a wide range of carrier densities is therefore essential to understand their coexistence.  

In this Letter, we reveal a universal hysteretic in-plane magnetoresistance in {\La}/{\Sr} heterostructures, characteristic of FM coexisting and competing with SC regardless of the carrier density or O$_2$ deficiency.  These results are consistent with the creation of pinned vortices by in-plane magnetic dipole moments, implying that FM is tightly confined to the interface, above a conventional 2D SC layer.  The existence of this confinement is confirmed by the observation of quantum oscillations from a non-magnetic electron gas deeper below the interface.

We have performed milliKelvin ac magnetotransport measurements on two distinct series of {\La}/{\Sr} heterostructures, A and B, grown by pulsed laser deposition (PLD).  A-type interfaces were grown at an O$_2$ pressure of 10$^{-3}$~mbar and subsequently annealed at 0.1~bar to reduce their O$^{2-}$ vacancy concentration, leading to two-dimensional (2D) carrier densities $n_{2D}~\sim~10^{13}$~cm$^{-2}$.  Hall bars patterned onto these heterostructures yield comparable results to the majority of SC {\La}/{\Sr} films in the literature~\cite{Reyren-2009,Bell-2009,Shalom-2010-2,Dikin-2011,Bert-2011}.  In contrast, B-type interfaces were also grown at 10$^{-3}$~mbar, but were not post-annealed: this was a deliberate ploy to maximise the O$^{2-}$ vacancy concentration - and hence $n_ {2D}$ - close to the interface, without creating a quasi-3D electron gas spanning the entire substrate~\cite{Herranz-2007}.  The resulting films exhibited $n_{2D}~\geq~10^{14}$~cm$^{-2}$, increasing to $\sim10^{15}$~cm$^{-2}$ upon field-effect doping.  Since all heterostructures behave similarly to the others in their series, we focus on two specific samples A and B for continuity, with SC critical temperatures $T_c$~=~0.28~K and 0.31~K respectively.  Further growth details, resistivity/Hall data and data-sets from additional samples may be found in the Supplementary Online Material (SOM).    

\begin{figure}[htbp]
\centering
\includegraphics [width=7.4cm,clip] {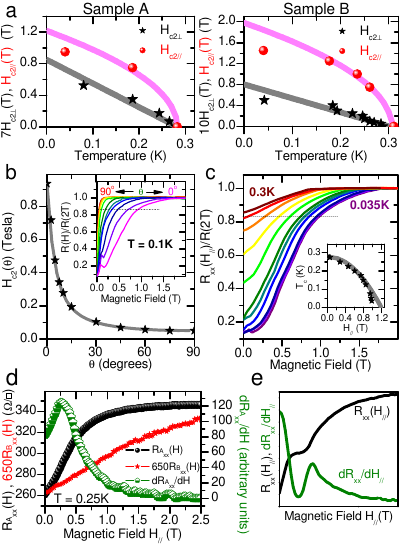}
\caption{\label{Fig1} (a) $H_{c2//,\perp}(T)$ for each sample, together with G-L fits (pink and grey lines; see SOM for the $R_{xx}(T)$ curves used to extract these data).  Small deviations from the fits are expected, since the G-L model only approximates $\xi(T)$ for $T{\ll}T_c$.  (b) $H_{c2}(\theta)$ at 0.1~K in sample A (black stars) and its predicted $H_{c2}(\theta)$ variation from the Tinkham model (grey line).  $\theta$=~0$^\circ$ and $\theta$=~90$^\circ$ correspond to $H_{//}$ and $H_\perp$ respectively.  Data are extracted from the normalised angular MR curves in the inset, where $H_{c2}(\theta,0.1~\mathrm{K})$ is defined by $R_{xx}(H_{c2})=R_{xx}(\mathrm{2T})-0.2(R_{xx}(\mathrm{2T})-R_{xx}(\mathrm{0T}))$ (dashed line).  (c) Temperature dependence of the normalised in-plane MR $R_{xx}(H_{//})/R_{xx}(\mathrm{2T})$ in sample A.  A similar 20\% fall in $R_{xx}$ is used to determine $T_c(H_{//})$ (inset, black stars).  $T_c(H_{//})$ is well fitted using a G-L model with $d$~=~18~nm.  (d) $R_{xx}(H_{//})$ at $T$~=~0.25~K for both samples.  The numerical derivative $dR_{xx}/dH_{//}$ is shown for film A, but $R_{xx}(H_{//})$ in film B is nearly 3 orders of magnitude smaller and too noisy to permit a similar treatment.  No demagnetization cycle was performed prior to these measurements and hence the peak in $dR_{xx}/dH_{//}$ is a weak MR signature of randomly-polarized FM zones.  (e) Qualitative sketch of the $R_{xx}(H_{//})$ expected in (d) from a helical SC.}
\end{figure}

\textit{Determining the depth of our SC channels:} In Fig.~\ref{Fig1}a, we fit the temperature-dependent perpendicular and parallel upper critical fields $H_{c2\perp,//}(T)$ using linearized Ginzburg-Landau (G-L) theory: $H_{c2\perp}(T)=\Phi_0/2\pi\xi^2(T)$ and $H_{c2//}(T)=\Phi_0\sqrt{3}/\pi\xi(T)d$, where $\Phi_0$ is the magnetic flux quantum.  Our fits yield G-L coherence lengths $\xi(0)=52\!\pm\!2$~nm, $60\!\pm\!2$~nm and SC channel widths $d=18\!\pm\!1$~nm, $9\!\pm\!1$~nm for A and B respectively.  $d\ll\xi$ and hence both films are 2D SC.  A scaling analysis~\cite{Reyren-2009} yields similar $d$, while back-gating film B with $V_g$~=~350~V increases $d$ to $19\!\pm\!2$~nm (SOM Figs.~S6b and S8b).  

\textit{Searching for unconventional SC:} Rashba spin-orbit coupling (RSOC) due to broken inversion symmetry at the interface also influences SC~\cite{Aoyama-2012} and may stabilise a helical FFLO state with maximal $T_c$ for in-plane fields $H_{//}>0$~\cite{Michaeli-2012}.  Since the RSOC is inversely proportional to $n_{2D}$~\cite{Shalom-2010-2}, we focus our search for exotic superconductivity on sample A.  Figure~\ref{Fig1}b shows the angle-dependent upper critical field $H_{c2}(\theta)$, accurately described by the Tinkham model for conventional SC films with $d\ll\xi$~\cite{Tinkham-1963}:
\begin{equation}
\label{Eq4}
\left|\frac{H_{c2}(\theta)\mathrm{sin}\theta}{H_{c2\perp}}\right|+\left(\frac{H_{c2}(\theta)\mathrm{cos}\theta}{H_{c2//}}\right)^2=1
\end{equation}
Even at $T$~=~0.1~K, $H_{c2//}$ exceeds the Pauli limit $H_P\equiv1.84~T_c$~=~0.52~T; this is consistent with the strong spin-orbit scattering expected at the interface.  Since $H_{c2//}>H_P$, the G-L method can only provide an upper limit for $d$; however, deviations between the calculated and true values of $d$ are small~\cite{Kim-2012} and have no impact on our discussion since SC remains confined within 20~nm of the interface.  In Fig.~\ref{Fig1}c, we track $T_c(H_{//})$ for sample A but find no peak at $H_{//}>0$; instead the data is closely reproduced by G-L theory.  To confirm this, we examine $R_{xx}(H_{//})$ at $T$=~0.25~K (just below $T_c(\mathrm{0T})$) for both samples in Fig.~\ref{Fig1}d: a maximum $T_c$ at $H_{//}>0$ will create a point of inflection or local minimum in $R_{xx}(H_{//})$, as depicted in Fig.~\ref{Fig1}e.  No such features are visible and we therefore find no support for a helical FFLO state~\cite{Note1}.  

\begin{figure}[htbp]
\centering
\includegraphics [width=7.1cm,clip] {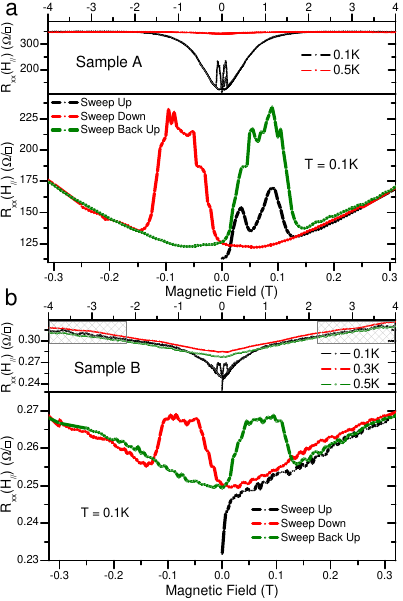}
\caption{\label{Fig2} (a) In-plane MR $R_{xx}(H_{//})$ for sample A ($n_{2D}$~=~2.3$\times$10$^{13}$~cm$^{-2}$, $T_c$~=~0.28~K) at temperatures 0.5~K and 0.1~K, with a low-field zoom on the $T$~=~0.1~K data (below).  These data were acquired after cooling from room temperature in zero field, resulting in randomly-oriented domains which generate the low peak structure seen in the sweep up.  (b) R$_{xx}(H_{//})$ for sample B at $n_{2D}$~=~2.4$\times$10$^{15}$~cm$^{-2}$ ($V_g$~=~350~V) for $T$~=~0.5~K, 0.3~K and 0.1~K, with a low-field zoom at $T$~=~0.1~K (below).  This loop was acquired after a separate measurement at $H_{//}>0$, leaving the FM domains polarized for the sweep up.  Shubnikov-de Haas oscillations emerge above 2.5~T (shaded boxes).}  
\end{figure}

\textit{Evidence for FM:} Figure~\ref{Fig2} displays in-plane MR loops for A and B.  Hysteresis is always observed at low fields ($\left|H\right|$~$<$~0.15~T), with peaks in $R_{xx}(H_{//})$ at negative/positive $H$ after sweeping down/up from large positive/negative $H$.  This pattern is distinct from the hysteresis seen in granular SC~\cite{Balaev-2007} and we hence absolve granularity from responsibility for our data.  In general, hysteretic MR is a signature of FM, but the disappearance of the hysteresis above $T_c$ indicates that SC also plays some role in its origin.  Non-zero resistance below $T_c$ for a SC in a magnetic field implies a loss of phase coherence, caused by mobile flux vortices.  Since the vortex diameter is $\sim\!2\xi$ and $d\ll\xi$, our SC channels are too shallow to allow vortex penetration parallel to an in-plane magnetic field: we must therefore identify a means of generating out-of-plane vortices using in-plane fields.  

\begin{figure}[tb]
\centering
\includegraphics [width=7cm,clip] {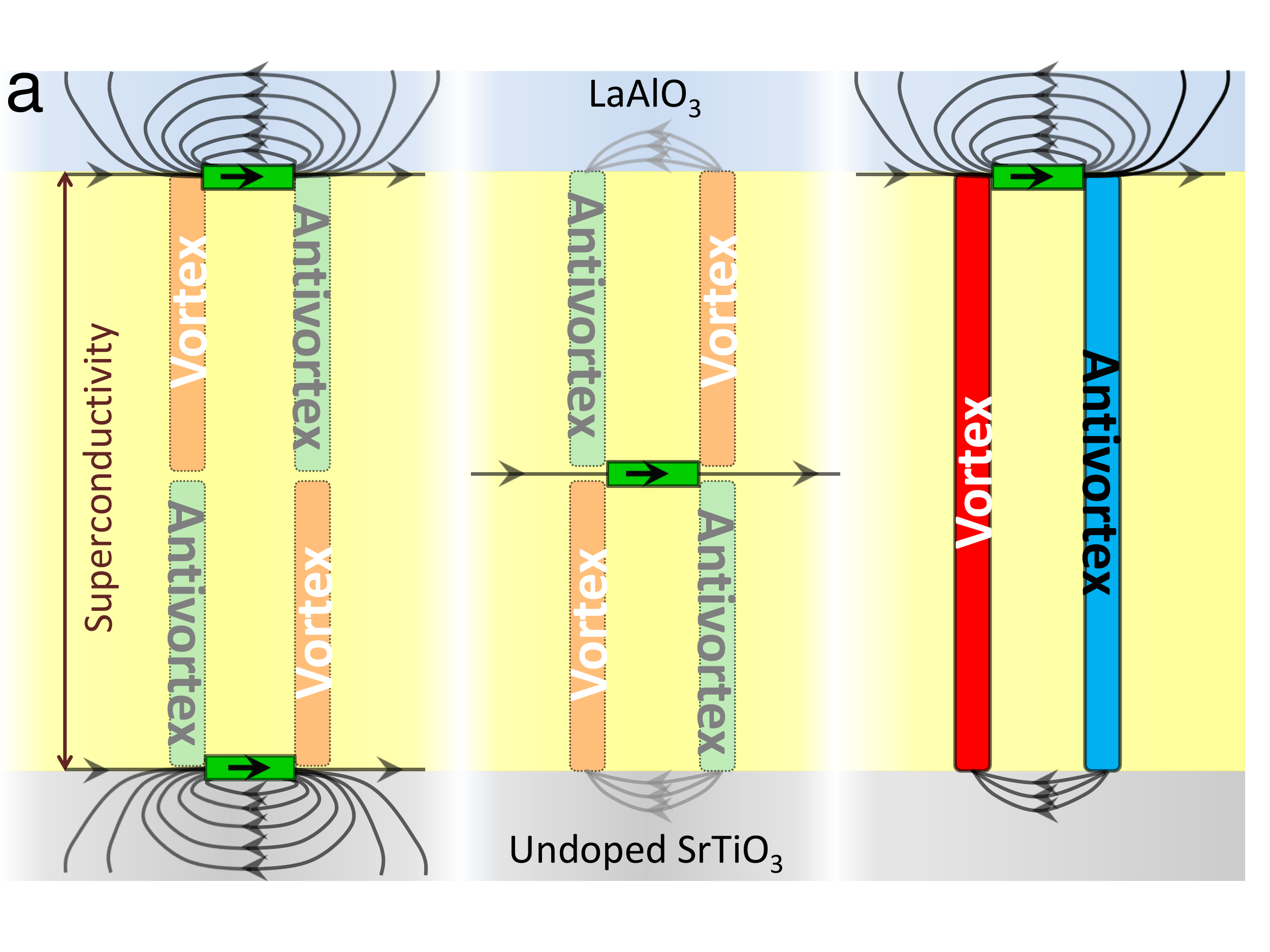}
\includegraphics [width=7cm,clip] {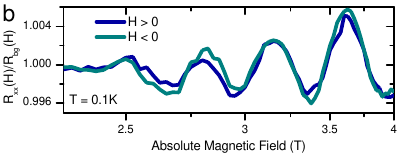}
\caption{\label{Fig3} (a) Out-of-plane vortex creation by in-plane FM dipoles.  We consider three configurations: a symmetric distribution of FM zones (green rectangles) above and below a SC channel (left); discrete FM zones buried within the channel (centre) and an asymmetric dipole distribution (right), with dipoles confined to either the top (shown) or the bottom of the channel.  (b) Comparison of the SdH oscillations in sample B upon reversing the direction of $H_{//}$ (data from shaded zones in Fig.~\ref{Fig2}b, normalised to a polynomial background $R_{bg}(H_{//})$.  The oscillation amplitude and phase are field-symmetric.}  
\end{figure}

\textit{Possible mechanisms for vortex creation:} Spatial inhomogeneities in the RSOC or the in-plane polarization can create vortices in helical superconductors~\cite{Aoyama-2012}; however such vortices would only appear above a critical field, which is incompatible with our low-field hysteresis.  It has been suggested that MR hysteresis in {\La}/{\Sr} arises from domain wall propagation in a continuous FM layer~\cite{Mehta-2012}, but this is implausible for several reasons.  Firstly, for thin FM layers the shape anisotropy energy is a key factor in domain formation and the walls should be of N{\'e}el rather than Bloch type, i.e. their magnetization points in-plane and so there is insufficient flux for out-of-plane vortex formation.  Secondly, there is no evidence in the literature for a continuous FM layer; in contrast, SQUID microscopy indicates an inhomogeneous FM distribution~\cite{Bert-2011}.  Thirdly, all our $R_{xx}(H)$ data are acquired at stable fields with our SC magnet in persistent mode, i.e. $dH/dt=0$ so there is no driving force for domain wall propagation (and our ac current negates any spin-torque transfer effect on the domains).  Therefore, the only realistic out-of-plane flux sources in {\La}/{\Sr} are the dipole fields from discrete FM zones, embedded in a SC channel and polarized in-plane (Fig.~\ref{Fig3}a).  To pass through the channel, out-of-plane components of the dipole fields must be quantised.  Arrays of vortices and antivortices therefore form around each pole, whose size and density depend on the geometry of each FM zone and its total dipole moment.  This situation is analogous to an artificial 2D SC/FM nanodot heterostructure~\cite{Milosevic-2004}.  

\textit{Where are the FM dipoles located?}  If the moments generating the vortices are buried deep in the SC channel (Fig.~\ref{Fig3}a centre panel), then no vortices will be generated, since the vortex/antivortex pairs at each end of the dipoles self-annihilate, leaving a purely horizontal field.  Also, while clearly revealing competition between FM and SC, the minor destructive effect of the polarized FM zones on SC (visible at $H_{//}=0$ in Figs.~\ref{Fig2}a,b;~S8a) does not support a large dipole population inside the channel.  If the moments are instead distributed symmetrically above and below the channel (left panel), then vortex/antivortex annihilation still tends to occur, with little quantised flux left in the channel.  Only an asymmetric FM zone distribution across the channel creates stable vortex/antivortex pairs (right panel), so FM must be confined to either the top or base of the channel.  

\textit{Is magnetism present below the SC channel, deep within {\Sr}?}  To address this possibility, we directly probe the electron gas in this region.  In sample B, Shubnikov-de Haas (SdH) oscillations develop for $H_{//}>$~2.5~T (Figs.~\ref{Fig2}b,~\ref{Fig3}b).  Since $d\leq$~20~nm and our field-effect doping proves that the {\Sr} is a good dielectric, we have not created a quasi-3D electron gas throughout the substrate~\cite{Herranz-2007}.  Nevertheless, the oscillations imply that a conducting ``tail'' extends into the {\Sr} over a distance at least twice the cyclotron radius $r_g={\hbar}k_F/eH$ (where $k_F$ is the Fermi wavevector).  At $n_{2D}=6.9\times10^{14}$~cm$^{-2}$, the SdH frequency $F$=~25~T: assuming a spherical Fermi surface for simplicity, $k_F=\sqrt{2{\pi}F/\Phi_0}$ (from the Onsager relation), $r_g$=~44~nm and hence the majority of the ``tail'' must lie below the base of the SC channel.  Now, SdH oscillations in magnetic materials should exhibit beating in the peak amplitude and field-reversal asymmetry~\cite{Abrikosov-1974}; moreover, scattering from localised electrons should suppress oscillations for $H<$~20-30~T~\cite{Lee-2011-2}.  From Fig.~\ref{Fig3}b, our oscillations are field-reversal symmetric, free from beating and emerge at merely 2.5~T; they hence originate from a non-magnetic electron gas.  Furthermore, the vertical field from any dipoles below the SC channel will be quantised, leading to a distorted flux profile when viewed from above: no evidence for this is seen by scanning SQUID~\cite{Bert-2011,Bert-2012,Kalisky-2012}.  We conclude that our {\Sr} is non-magnetic and FM is strictly confined to the interface.   

\begin{figure}[htbp]
\centering
\includegraphics [width=6.6cm,clip] {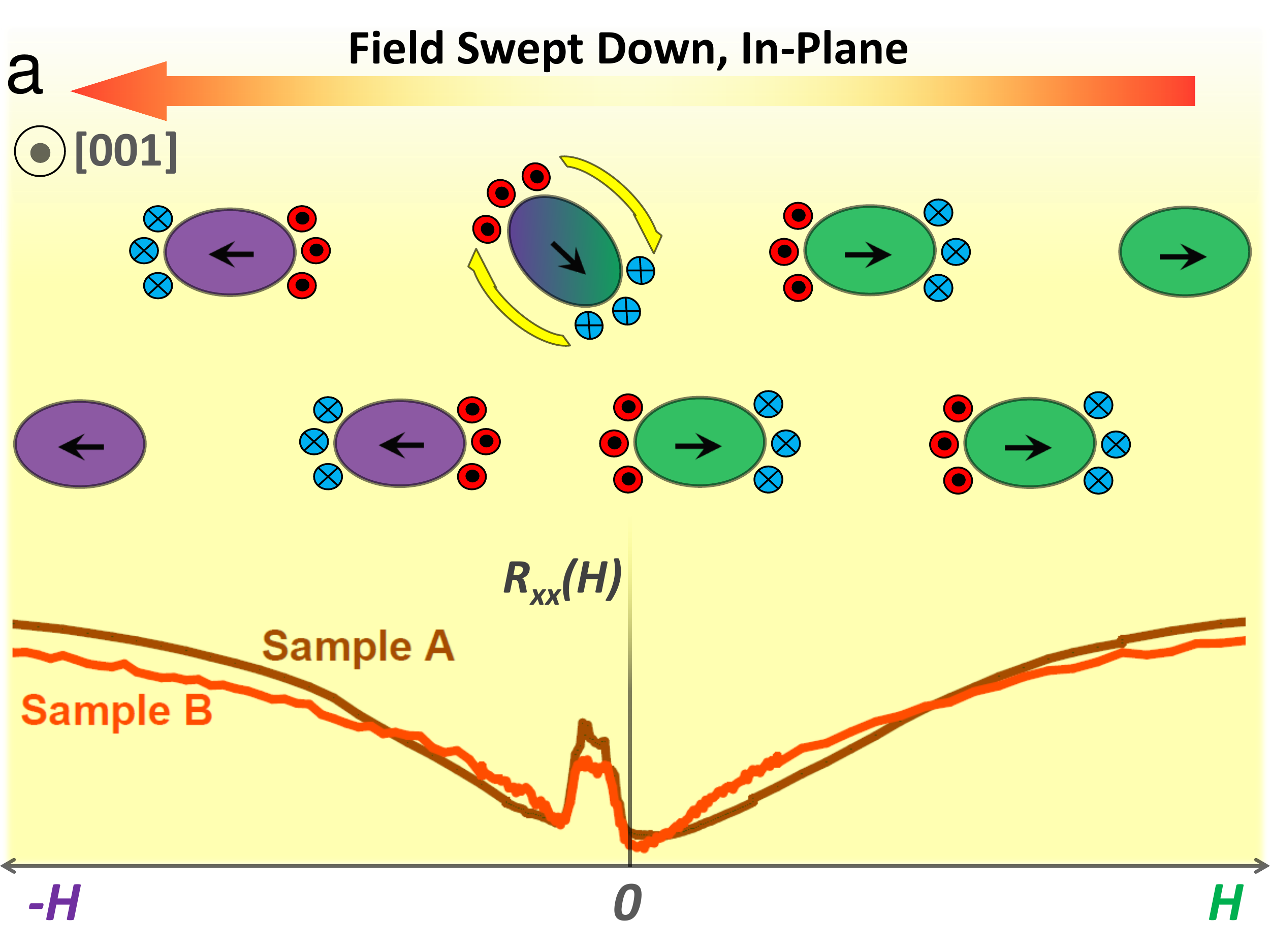}
\includegraphics[width=7.2cm,clip] {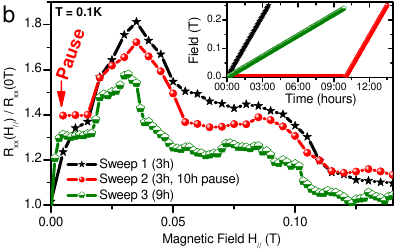}
\caption{\label{Fig4} (a) Repolarization mechanism for FM domains upon sweeping from $H_{//}\gg0$ to $H_{//}\ll0$.  Green and purple denote opposing polarizations, while vortices are red and antivortices blue.  Yellow arrows indicate the rotation of the domains at fields $\left|H\right|>H_{coerc}$, dragging the vortex/antivortex pairs with them.  (b) Amplitude variation of the hysteretic peak in the normalised magnetoresistance $R_{xx}(H_{//})/R_{xx}(0\mathrm{T})$ for 3 distinct time-dependent field sweeps (see text for details; data from an A-type heterostructure).  Sweeps were performed directly after polarization at $H$~=~-4~T.  Each data-point was acquired in a stable, constant field, 90~s after our SC magnet entered persistent mode ($dH/dt=950\mu$Ts$^{-1}$ while ramping the field) and our ac current flowed throughout the experiment.  The field increment between data-points determines the total measurement duration (inset) and hence the peak amplitude.  The shape of the peak (corresponding to the pinning profile in the channel) is stable between sweeps, but exhibits small changes after thermal cycling to 300~K.} 
\end{figure}

\textit{How do the vortices generate a hysteretic MR?}  Peak formation in $R_{xx}(H_{//})$ corresponds to reversing the FM polarization, a process which is illustrated in Fig.~\ref{Fig4}a.  This will mainly occur via dipole rotation rather than domain wall propagation, due to the limited zone size ($\lesssim$~10~$\mu$m from SQUID data), the presence of hysteresis up to $\left|H\right|$=~0.15~T (implying a large coercive field) and flux conservation within the channel.  During repolarization, a torque $\mathbf{m}\times\mathbf{H}$ is exerted on the FM zones, where $\mathbf{m}$ is the dipole moment.  The vortex/antivortex arrays around the poles are dragged through a 180$^\circ$ rotation, creating an electric field in their cores and dissipating energy.  However, vortex motion is impeded by pinning, hence broadening the MR peak.  Within the framework of the Anderson flux creep model~\cite{Anderson-1962}, 
\begin{equation}
\label{Eq2}
\nu_{depin}~{\propto}~e^{-(U-\left|m\right|\left|H\right|+\mathbf{m}\cdot\mathbf{H})/k_BT}
\end{equation}
where $\nu_{depin}$ is the vortex depinning frequency, $U$ is the vortex pinning potential, $\left|H\right|$ exceeds the FM coercive field $H_{coerc}$ and we disregard the effects of our ac current.  $\nu_{depin}$ scales linearly with the induced electric field and hence $R_{xx}(H_{//})$.  Assuming an infinite vortex supply, constant $U$ and domain polarization antiparallel to $H$, an increase ${\Delta}H$ will generate ${\Delta}R_{xx}(H_{//}){\propto}e^{2mH}{\Delta}H$.  These assumptions are clearly unrealistic, since our samples contain finite numbers of vortex-inducing FM dipoles which may become pinned at arbitrary angles to the field during repolarization; we also expect a broad variation in the local pinning potential within the SC channel from the inhomogeneous defect distribution.  Nevertheless, this relation permits a qualitative understanding of the evolution of our hysteretic peaks over time (Fig.~\ref{Fig4}b).  Compared with a reference data-set (sweep 1) acquired at field increment $\delta_H$, a long pause in the data acquisition at $H$~=~0.005~T (sweep 2) has little effect on $R_{xx}(H_{//})$, since $H_{coerc}>$~0.005~T for most of the domains and depinning is limited.  In contrast, measuring with no pause but a smaller field increment $\delta_H/3$ (sweep 3) results in a larger drop in $R_{xx}(H_{//})$, since fewer vortices are depinned at each data point.  Above $H_{//}\sim$~0.15~T, all FM domains have been repolarised (i.e. all vortex/antivortex arrays have been rotated through 180$^\circ$) and our ac current is the only remaining depinning force: $R_{xx}(H_{//})$ therefore falls to its background level.

\textit{A picture emerges of a narrow layer of FM clusters confined to the interface, above a 2D SC layer.}  The inhomogeneity in the FM and its independence from $n_{2D}$ suggest that it originates from a static defect distribution, in agreement with several recent models~\cite{Pavlenko-2012,*Pavlenko-2012-2,Banerjee-2013}.  Previously, signatures of FM have been linked to high O$_2$ pressure growth~\cite{Brinkman-2007,Ariando-2011}, but this is due to enhanced defect-induced localisation in the 2D limit and FM transport signatures being ``short-circuited'' by mobile electrons below the interface at higher $n_{2D}$.  Together with data from such resistive interfaces, our work shows that FM can exist across the entire {\La}/{\Sr} phase diagram: $10^{12}{\leq}n_{2D}{\leq}10^{15}$~cm$^{-2}$.  The similarities in the hysteretic MR of our two film series indicate that O$^{2-}$ vacancies are neither essential nor anathemic to FM; however scattering from vacancies and cation intermixing may help to stabilise FM.  We note that beyond the hysteretic in-plane MR and Kondo effect in $R_{xx}(T)$ at low temperature (see SOM), there is no obvious evidence for magnetism in our heterostructures.  It is therefore likely that other {\La}/{\Sr} films studied in the literature contained FM inclusions, but their presence was missed since parallel MR loops were not acquired below $T_c$.   

In conclusion, our work illustrates how the reduced symmetry, modified electronic structure and defects inherent to an interface promote an alien emergent phase (FM) to interact and compete with the usual emergent ground state in doped {\Sr} (SC).  Although competition only occurs at the top of the SC channel, the asymmetric distribution of the FM influences the entire SC layer by generating vortices even at zero applied field.  Looking to the future, we propose that atomically precise defect control may enable the development of hybrid FM/SC devices in {\La}/{\Sr} films, such as spintronic latches, vortex-based memory or SC qubits.

The authors gratefully acknowledge discussions with I. Martin, W. Pickett and J. Chakhalian.  This work was supported by the National Research Foundation, Singapore, through Grant NRF-CRP4-2008-04.  The research at the University of Nebraska-Lincoln (UNL) was supported by the National Science Foundation through the Materials Research Science and Engineering Center (Grant DMR-0820521) and the Experimental Program to Stimulate Competitive Research (Grant EPS-1010674).

\section{Supplementary Material}
\subsection*{1. Synthesis and Measurement Techniques}

Our two series of heterostructures A and B were both grown using a standard pulsed laser deposition system manufactured by Twente Solid State Technology B.V., equipped with a reflection high-energy electron diffraction (RHEED) facility.  10 unit cells of {\La} were deposited on TiO$_2$-terminated 0.5mm thick commercial {\Sr} (001) substrates from Shinkosha.  The total incident laser energy was 9~mJ, focussed onto a 6~mm$^2$ rectangular spot.  For both samples, the oxygen pressure was maintained at 10$^{-3}$~mbar during growth at 800$^\circ$C.  However, series A underwent a subsequent annealing stage: after cooling to 500$^\circ$C at 10$^{-3}$~mbar, the oxygen pressure was increased to 0.1~bar.  The temperature was held at 500$^\circ$C for 30 minutes before natural cooling to room temperature, still in 0.1~bar oxygen.  In contrast, series B was allowed to cool naturally to room temperature in 10$^{-3}$~mbar oxygen.  

Hall bars with Au-Ti contact pads were patterned onto the top surface of the {\La} using electron-beam lithography, while sample B also had an Au-Ti back gate sputter-deposited across the entire base of the {\Sr} substrate prior to patterning.  The Hall bar width was 80~$\mu$m and the voltage contact separation 660~$\mu$m.  Patterning onto the {\La} surface rather than directly contacting the interface allows us to remain sensitive to AlO$_2$ surface transport in parallel with interfacial states, even when the interface is superconducting (see section 2).  

Transport measurements were performed in a Janis cryogen-free dilution refrigerator, using an ac technique with two SRS 830 lock-in amplifiers and a Keithley 6221 AC current source.  All data were acquired using an excitation current of 500~nA at 19~Hz; this value was chosen to maximise the signal-to-noise ratio whilst eliminating any sample heating effects at milliKelvin temperatures.  Our noise threshold is approximately 2~nV.  The evolution of the capacitance ($\sim$~1~nF) of the {\Sr} substrate with gate voltage $V_g$ was verified using a General Radio 1621 manual capacitance bridge: no leakage or breakdown occurred even at $V_g$~=~500~V.  

\subsection*{2. Electrical Resistivity and Superconducting Transitions}

\begin{figure}[htbp]
\centering
\includegraphics*[width=8cm,clip]{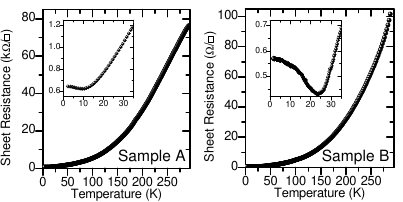}
\caption[]{\label{FigS0} Variation of the sheet resistance $R_{xx}(T)$ upon cooling from room temperature for sample A (left) and B (right).  The resistance below 35~K is magnified in the insets to each graph, illustrating the Kondo minima.
}
\end{figure}

Upon cooling from room temperature, the electrical resistance of both heterostructures passes through a minimum at low temperature (fig.~S\ref{FigS0}), then rises logarithmically prior to the onset of superconductivity at $\sim$~0.3~K.  The minima are located at $T_m$~=~10~K and 25~K for samples A and B respectively.  A logarithmic divergence in the low-temperature resistance is a signature of the Kondo effect, i.e. scattering off dilute magnetic impurities.  Within the Kondo scenario, sample B (which has a high oxygen vacancy concentration) must contain a higher density of magnetic scattering centres due to its higher $T_m$.   
  
\begin{figure}[htbp]
\centering
\includegraphics*[width=7.5cm,clip]{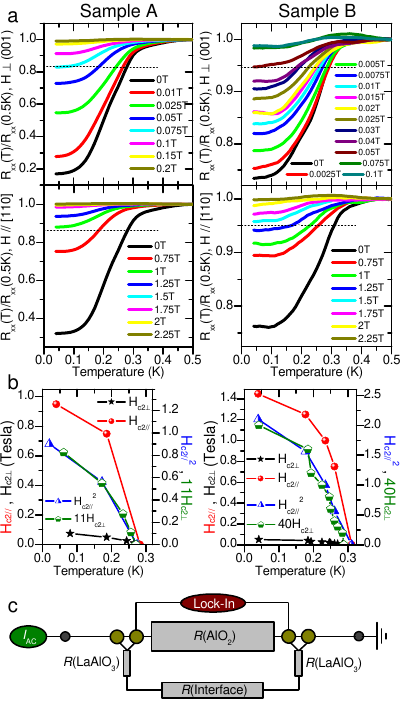}
\caption[]{\label{FigS1} (a) Temperature dependence of the longitudinal resistance $R_{xx}(T)$ in sample A (left) and B (right) within the superconducting (SC) phase, for a range of magnetic fields applied perpendicular to the interface  ($H{\perp}(001)$, above) and parallel to the interface ($H//[110]$, below).  All data are normalised to the resistance at $T$~=~0.5~K.  (b) Temperature dependence of the parallel and perpendicular upper critical fields $H_{c2//,\perp}(T)$ for each heterostructure.  These data are extracted from our $R_{xx}(T)$ curves: we define $T_c(H){\equiv}H_{c2}(T)$ as the temperatures at which $R_{xx}$ falls by 20\% of the difference between its values at 0.5~K and 0.04~K in zero field, indicated by the dashed lines intersecting the two $R_{xx}(T)$ plots in (a).  Note that this 20\% criterion is arbitrary, since although changing the percentage will lead to small variations in the calculated coherence length $\xi$, the anisotropy and hence our determination of the superconducting channel thickness $d$ will remain unchanged.  A scaling analysis for 2D SC~\cite{SSchneider-2008,SReyren-2009} is also shown, which provides an alternative means to determine $d$ using $H^{2}_{c2//}(T)$~=~$\frac{\pi\Phi_0}{2d^2}$$H_{c2\perp}(T)$.  This yields SC layer thicknesses $d=16\!\pm\!1$~nm and $9\!\pm\!1$~nm for samples A and B respectively, in excellent agreement with the values from Ginzburg-Landau theory ($18\!\pm\!1$~nm and $9\!\pm\!1$~nm).  (c) Schematic of the experimental setup and Hall bar patterning, indicating the resistive component from the 10 unit cells of {\La} which we always measure in series with the superconducting interface, together with the AlO$_2$ surface states in parallel with the interface.
}
\end{figure}

The resistive transitions to the superconducting state used to determine the critical fields in fig.~1 of our manuscript are shown in fig.~S\ref{FigS1}a.  Although the transitions qualitatively follow the behaviour expected for a quasi-2D superconducting film, the resistance does not fall to zero even at the lowest temperatures measured (0.035~K).  This does not necessarily imply that our interface is inhomogeneous; in fact, non-zero resistance is a natural consequence of the patterning technique which we have used.  A schematic diagram for our pattern is shown in fig.~S\ref{FigS1}c: since the contacts for our Hall bars are only deposited onto the top surface of the {\La}, there is no direct contact with the conducting interface.  Instead, contact is made vertically through the 10 unit cells of {\La}, which exhibits a weak conductivity dependent on the oxygen vacancy concentration.  This provides a small resistive component in series with the interface, leading to a measured non-zero resistance even with a homogeneous superconducting interface.  Any conducting AlO$_2$ surface states will generate a parallel contribution to the measured resistance; the advantage of this patterning technique is that it enables these surface states to be probed without being ``shorted out'' by the superconducting interface.



Conversely, this does not easily permit us to differentiate between homogeneous and inhomogeneous superconducting layers (or even field-induced inhomogeneous nucleation of superconductivity).  However, this does not affect the conclusions of our work in any manner: firstly, inhomogeneities at the interface are entirely expected due to the tendency of the $d_{xy}$ electrons to localise and form ferromagnetic zones where superconductivity is suppressed.  For sufficiently high densities of ferromagnetic inclusions above a shallow superconducting channel, the percolative zero-resistance current path may vanish.  Secondly, the observed quantum oscillations originate from carriers deeper below the interface and only emerge at high magnetic fields when the superconductivity has been entirely quenched.  Thirdly, if we consider the oxygen-deficient high carrier density heterostructure B, it is plausible that clusters of oxygen vacancies at the AlO$_2$ surface could locally overdope the interface beyond its maximum superconducting carrier density within certain discrete zones (also enhancing the ferromagnetism, as suggested by our Kondo effect and theory~\cite{SPavlenko-2012,*SPavlenko-2012-2}).  This is a natural consequence of vacancy-doping the {\La}/{\Sr} interface and in no way reflects negatively on our results.  

Furthermore, using data for the critical current density from the literature~\cite{SStornaiuolo-2012} ($\sim$~40~nA per micron channel width), we estimate that the critical current in our Hall bars is of the order of several microAmps at zero field.  Our measurement current (500~nA) is only one order of magnitude smaller than this value, thus contributing to the broadening of the transitions which we see in a magnetic field.  We stress that a current of this magnitude is essential to achieve an adequate signal-to-noise ratio in highly conductive materials such as sample B, especially for Shubnikov-de Haas measurements.  It also facilitates depinning by exciting a lateral ``shaking'' force on vortices, thus influencing (though not causing) the hysteresis in our in-plane magnetoresistance data.  

\subsection*{3. Hall Effect Data}

\begin{figure}[htbp]
\centering
\includegraphics*[width=8cm,clip]{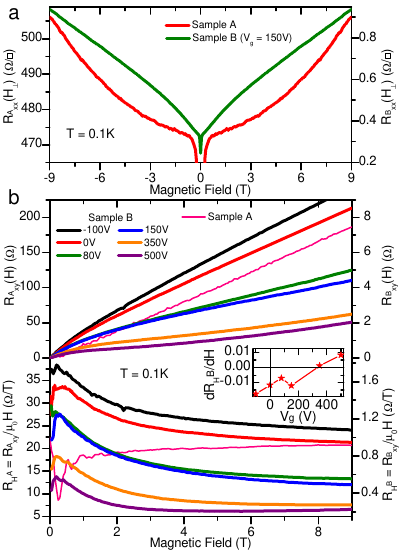}
\caption[]{\label{FigS2} (a) Perpendicular magnetoresistance $R_{xx}(H_\perp)$ for both heterostructures.  (b) Above: Hall resistance $R_{xy}$ for both samples, including gate voltage dependence for sample B.  Below: Field-dependent Hall coefficient $R_H$ for each sample, highlighting the Lorentzian form for sample B characteristic of a multi-band conductor.  Inset: variation of the gradient of the Hall coefficient $dR_H/dH$ with gate voltage at $H$~=~9~T.  
}
\end{figure}

All stated sheet carrier densities in our work have been obtained by linear fits to the high-field Hall resistance $R_{xy}(H)$ ($H>$~5~T), where $n_{2D}\frac{dR_{xy}}{dH}=-\frac{1}{e}$.  However, both our Shubnikov-de Haas effect and various magnetotransport studies in the literature~\cite{SBell-2009,SShalom-2010,SJoshua-2011} have revealed evidence for multiple conduction bands at the interface, which should lead to Lorentzian forms for both the perpendicular magnetoresistance (MR) $R_{xx}(H_{\perp})$ and the Hall coefficient $R_H$.  We plot $R_{xx}(H_{\perp})$ and $R_H$ in fig.~S\ref{FigS2}.  

The simple message which we wish to convey here is the dramatically different behaviour of both $R_{xx}$ (fig.~S\ref{FigS2}a) and $R_H$ (fig.~S\ref{FigS2}b) for the two heterostructures.  Examining the MR first, we observe that the curvature of $R_{xx}$ is positive for sample A, compared with negative curvature and a Lorentzian form in sample B.  The Hall effect is even more revealing, with $R_H$ in sample A displaying approximately linear behaviour, which for {\La}/{\Sr} implies $d_{xy}$ single-band occupancy (slight deviations from linearity may be due to limited hole-like contributions from carriers at the AlO$_2$ top surface).  In contrast, $R_H$ in sample B again shows the Lorentzian shape expected for a two-band system (i.e. $d_{xy}$ and $d_{xz,yz}$ band occupancy).  Beyond the ability to extract the total carrier density, the failings of simple two-band Hall coefficient models are well-known~\cite{SJoshua-2011} due to the significant differences in the field dependence of the mobilities of carriers in each band; we therefore do not attempt any more detailed quantitative analysis of our data.  One further unusual feature in our Hall data is worthy of mention: a crossover in the gradient of $R_H$ from negative to positive at high gate voltages (fig.~S\ref{FigS2}b, inset).  This cannot be explained by a simple interfacial two-band model and is a consequence of the gradual population of states deeper within the substrate.  

We note that the total carrier density which we measure for sample A is 2.3$\times$10$^{13}$~cm$^{-2}$, slightly larger than the critical density $1.68\!\pm\!0.18\!\times\!10^{13}$~cm$^{-2}$ recently obtained for the Lifshitz transition~\cite{SJoshua-2011}.  However, we see no evidence for two-band behaviour in the Hall coefficient of sample A and we conclude that the extra carriers which we detect most probably originate from deeper within the {\Sr} substrate (since any AlO$_2$ surface states should still be hole-like within this doping range).  This is an important point, since it absolves the $d_{xz,yz}$ bands of responsibility for generating our observed ferromagnetic domains.  

\subsection*{4. Coexistent Ferromagnetism and Superconductivity, and their Evolution with Gate Voltage}

\begin{figure}[htbp]
\centering
\includegraphics*[width=8cm,clip]{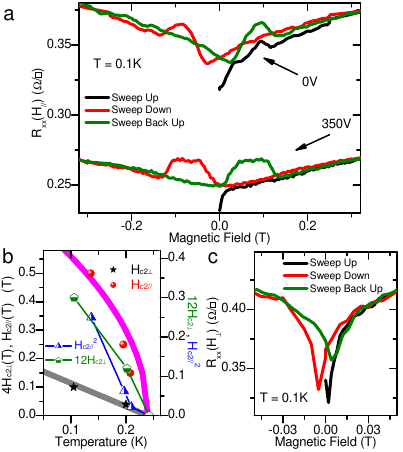}
\caption[]{\label{FigS3} (a) Comparison of the hysteretic parallel magnetoresistance $R_{xx}(H_{//}$ at $V_g$~=~0~V and 350~V.  (b) Parallel and perpendicular upper critical fields $H_{c2//,\perp}$ at $V_g$~=~350~V, together with Ginzburg-Landau fits (pink and grey lines) indicating a coherence length $\xi$~=~82~$\pm$~2~nm and SC channel thickness $d~$~=~19~$\pm$~2~nm.  A 2D scaling analysis which yields $d$~=~16~$\pm$~3~nm is also shown.  (c) Hysteresis in the perpendicular magnetoresistance $R_{xx}(H_\perp)$.  The minima in $R_{xx}$ occur at $\pm$~0.005~T, which is of the same order of magnitude as the remanent field in most large superconducting coils.
}
\end{figure}

Let us consider the effects of applying a gate voltage to sample B on the ferromagnetic and superconducting phases.  SQUID microscopy studies have indicated that gating the {\La}/{\Sr} interface to modulate its carrier density does not have any effect on the density of ferromagnetic inclusions~\cite{SKalisky-2012,SBert-2012}.  In contrast, applying a positive gate voltage to increase $n_{2D}$ in sample B does change the shape of the hysteretic peaks, which are broader and clearer at $V_g$~=~350~V (fig.~S\ref{FigS3}a).  

Numerous potential explanations exist for this effect.  Firstly, we suggest that the electric field across the {\Sr} may lead to a further increase in the vortex mobility once depinning has occurred, thus increasing the measured resistance even for small applied fields.  We must also consider the expansion of the superconducting channel upon gating: the channel roughly doubles in thickness between $V_g$~=~0~V and 350~V (fig.~S\ref{FigS3}b).  A thicker channel will increase the probability of pinning any given vortex during the rotation of its respective ferromagnetic dipole, hence broadening the hysteretic peak.  However, the ``hardest'' pins (from the largest defects) will be located closer to the interface and hence the maximum field at which hysteresis is observed should remain similar: from fig.~S\ref{FigS3}a, this is indeed the case.  Another relevant factor in modifying the hysteretic peak shape may be a partial suppression of superconductivity very close to the interface due to the high carrier density in sample B.  This would also explain its narrower as-grown superconducting channel compared with sample A, although it is important to remember that the as-grown vertical charge distribution profile is a crucial factor in determining the absolute superconducting layer thickness and this may vary significantly between heterostructures.  

A very small hysteresis is visible in the out-of-plane MR $R_{xx}(H_\perp)$, although this occurs at fields close to zero, comparable to the typical remanence in superconducting magnets.  For completeness, we plot this in fig.~S\ref{FigS3}c.  We stress that the absence of any large hysteretic peaks in $R_{xx}(H_\perp)$ is entirely expected, since in this configuration there is no rotation of the in-plane moments; instead, the vortex density is much higher and we enter a liquid phase at very low applied fields.

\subsection*{5. Data Reproducibility}
\begin{figure*}[htbp]
\centering
\includegraphics*[width=17.5cm,clip]{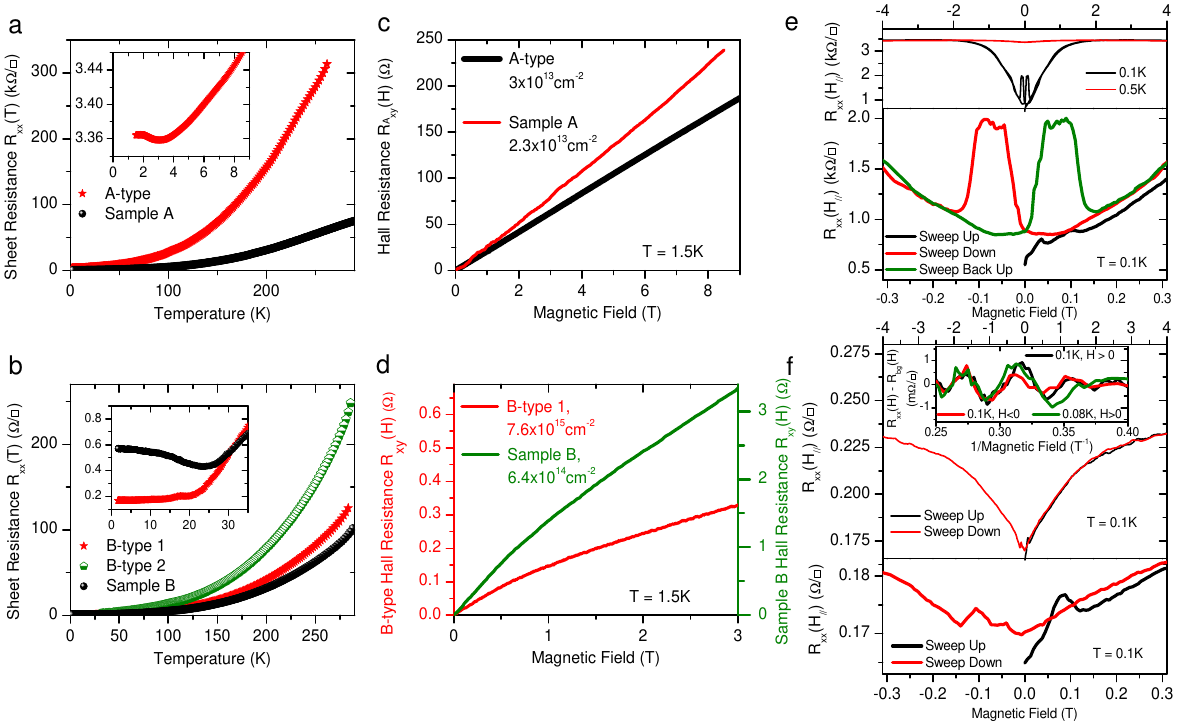}
\caption[]{\label{FigS4} (a) Temperature-dependent resistance R(T) of another A-type {\La}/{\Sr} heterostructure, grown by pulsed-laser deposition (PLD) at 10$^{-3}$~mbar and post-annealed in 0.1~mbar O$_2$. Data from sample A studied in our manuscript is included for comparison.  Inset: Kondo effect in this A-type heterostructure.  \newline(b) R(T) in two B-type {\La}/{\Sr} heterostructures (grown by PLD at 10$^{-3}$~mbar with no post-annealing), compared with sample B from our manuscript.  The low temperature resistance is less than 1$\Omega/\square$ in all three heterostructures.  Inset: zoom on R(T) at low temperature for B-type heterostructure 1, with data from sample B for comparison.  Although the Kondo effect is not as clear in this film as in sample B (due to a large parallel conductance from mobile electrons deeper within the non-magnetic bulk {\Sr}), a distinct kink is still present below 25~K and is indicative of a high density of magnetic scattering centres.  \newline(c) Hall effect at 1.5~K for the A-type heterostructure shown in (a), together with data from sample A in our manuscript.  Both heterostructures exhibit single-band transport, with 2D carrier densities $n_{2D}$ = 2.3$\times$10$^{13}$~cm$^{-2}$ for sample A and 3.0$\times$10$^{13}$~cm$^{-2}$ for the second A-type heterostructure.  It should be noted that the low-temperature values of R(T) (roughly 1k$\Omega/\square$) and $n_{2D}$ in our A-type heterostructures are typical of annealed {\La}/{\Sr} films in the literature.  \newline(d) Hall effect at 1.5~K for the B-type heterostructure 1 shown in (b), together with data from sample B in our manuscript.  The Hall resistance exhibits a similar non-linear trend in both heterostructures, which is characteristic of multi-band transport.  We measure $n_{2D}$~=~7.6$\times$10$^{15}$~cm$^{-2}$ for the B-type heterostructure 1, compared with 6.4$\times$10$^{14}$~cm$^{-2}$ for sample B.  \newline(e) Interfacial ferromagnetism and superconductivity in the 3$\times$10$^{13}$~cm$^{-2}$ A-type heterostructure from figs.~(a,c).  Hysteretic behaviour for H~$<$~0.16~T is observed in the in-plane magnetoresistance below the superconducting transition, similar to that seen in sample A (Fig. 2a in our manuscript).  \newline(f) Interfacial ferromagnetism and superconductivity in the 7.6$\times$10$^{15}$~cm$^{-2}$ B-type heterostructure from figs.~(b,d).  The in-plane magnetoresistance is hysteretic for H~$<$~0.15T, although the amplitude of the hysteresis is slightly smaller than that in sample B (fig.~2b in our manuscript) due to the larger parallel conductance from the high-mobility electron gas deeper within the {\Sr} substrate.  Shubnikov-de Haas oscillations are visible at in-plane fields above 2.5~T (top inset), with a characteristic frequency of 23.8~T, similar to that reported in sample B.  
}
\end{figure*}
Over an 18-month period spanning the duration of this project, we synthesized numerous ``A-type'' and ``B-type'' heterostructures, all of which exhibited qualitatively similar behaviour.  Series A have $n_{2D}\sim10^{13}$~cm$^{-2}$ and display single-band transport, while $n_{2D}>10^{14}$~cm$^{-2}$ and the Hall coefficient is non-linear in series B.  Superconductivity is observed to coexist with ferromagnetism regardless of the carrier density, while Shubnikov-de Haas oscillations emerge in parallel fields below series B interfaces. Data-sets from several other heterostructures may be found in fig.~S\ref{FigS4}.

\end{document}